\documentclass[runningheads]{llncs}
\usepackage{times}
\usepackage{latexsym}
\usepackage{url}
\usepackage{graphicx}
\usepackage{dblfloatfix}
\usepackage{amssymb}
\usepackage{multirow}
\usepackage{amsmath}
\usepackage{tabulary}
\usepackage{booktabs}
\usepackage{makecell}
\usepackage[dvipsnames]{xcolor}
\usepackage{comment}
\usepackage{caption}
\usepackage{xcolor}
\usepackage{hyperref}

\begin{document}
%
\title{Improving Scholarly Knowledge Representation:\\ Evaluating BERT-based Models for \\ Scientific Relation Classification 
}
%
\titlerunning{Evaluating BERT-based Models for SR Classification}
%
\author{Ming Jiang\inst{1}\orcidID{} \and
Jennifer D’Souza\inst{2}\orcidID{0000-0002-6616-9509} \and \\ 
S\"{o}ren Auer\inst{2}\orcidID{0000-0002-0698-2864} \and
J. Stephen Downie\inst{1}\orcidID{0000-0001-9784-5090}}
\authorrunning{M. Jiang et al.}
%
\institute{University of Illinois at Urbana Champaign, USA \and
TIB Leibniz Information Centre for Science and Technology and \\L3S Research Center at Leibniz University of Hannover, Hannover, Germany \\
\email{\{mjiang17|jdownie\}@illinois.edu, \{jennifer.dsouza|auer\}@tib.eu}}

%
\maketitle              
\begin{abstract}
With the rapid growth of research publications, there is a vast amount of scholarly knowledge that needs to be organized in digital libraries. To deal with this challenge, techniques relying on knowledge-graph structures are being advocated. Within such graph-based pipelines, inferring relation types between  related scientific concepts is a crucial step. Recently, advanced techniques relying on language models pre-trained on the large corpus have been popularly explored for automatic relation classification. Despite remarkable contributions that have been made, many of these methods were evaluated under different scenarios, which limits their comparability. To this end, we presents a thorough empirical evaluation on eight \textsc{Bert}-based classification models by focusing on two key factors: 1) \textsc{Bert} model variants, and 2) classification strategies. Experiments on three corpora show that domain-specific pre-training corpus benefits the \textsc{Bert}-based classification model to identify the type of scientific relations. Although the strategy of predicting a single relation each time achieves a higher classification accuracy than the strategy of identifying multiple relation types simultaneously in general, the latter strategy demonstrates a more consistent performance in the corpus with either a large or small size of annotations. Our study aims to offer recommendations to the stakeholders of digital libraries for selecting the appropriate technique to build knowledge-graph-based systems for enhanced scholarly information organization.

\end{abstract}

\section{Introduction}
Today scientific endeavors are increasingly facing a publication deluge~\cite{stm}, which results in the rapid growth of document-based scholarly publications in digital libraries. While abundant resources of scholarly information have been provided in digital libraries, it is still challenging for researchers to obtain a comprehensive, fine-grained and context-sensitive scholarly knowledge for their research---a problem that is more acute in multi-disciplinary research~\cite{Jaradeh2019ORKG}. According to \cite{auer2018towards,auer2019orkg}, current keyword-based methods for indexing scholarly articles may not be able to cover all aspects of knowledge involved in each article. Further, the single keyword search on scholarly articles fails to consider the semantic associations among the units of scholarly information. Thus, toward better scholarly knowledge organization in digital libraries, some initiatives~\cite{Jaradeh2019ORKG,stephen} advocate for building an interlinked and semantically rich knowledge graph structure combining human curation with machine learning.

A key to build a knowledge graph from a scholarly article is to identify relations between scientific terms in the article. In the natural language processing (NLP) community, within the context of human annotations on the abstracts of scholarly articles~\cite{augenstein2017semeval,gabor2018semeval}, seven relation types between scientific terms have been studied. They are \textsc{Hyponym-Of}, \textsc{Part-Of}, \textsc{Usage}, \textsc{Compare}, \textsc{Conjunction}, \textsc{Feature-Of}, and \textsc{Result}. The annotations are in the form of generalized relation triples: $\langle$experiment$\rangle$ \textsc{Compare} $\langle$another experiment$\rangle$; $\langle$method$\rangle$ \textsc{Usage} $\langle$data$\rangle$; $\langle$method$\rangle$ \textsc{Usage} $\langle$research task$\rangle$. Since human language exhibits the paraphrasing phenomenon, identifying each specific relationship between scientific concepts is impractical. In the framework of an automated pipeline for generating knowledge graphs over massive volumes of scholarly records, the task of classifying scientific relations (i.e., identify the appropriate relation type for each related concept pair from a set of predefined relations) is therefore indispensable. 

In this age of the ``deep learning tsunami'', many studies have developed neural network models to improve the construction of automated scientific relation (SR) classification systems~\cite{manning2015computational}. With the recent introduction of language pre-training techniques such as \textsc{Bert}~\cite{bert} models, the opportunity to obtain boosted machine learning systems is further accentuated. While prior work~\cite{scibert,mre19} has demonstrated high classifier performances, the evaluation of these studies were mainly conducted under a single scenario, e.g. the testing data is from a single resource. This may leads to the difficulty of obtaining comparable results and conclusive insights about the robustness of the classifiers in practice. For example, academic digital libraries are hard to select the appropriate technique to improve their knowledge organization services based on their specific conditions such as the collection scale and diversity.

To help to fill in the aforementioned gap, we conducted an empirical evaluation on the advanced techniques using pre-trained language models for SR classification. In particular, we implemented and analyzed eight \textsc{Bert}-based classification models by exploring the impact of two key factors: 1) classification strategies (i.e., predicting either a single relation or multiple relations at one time); and 2) \textsc{Bert} model variants with respect to the domain and vocabulary case of the pre-training corpus. To further explore the potential influence of data settings, we assess the performance of each model on three corpora including: 1) a single-domain corpus with sparse relation annotations on scholarly publication abstracts in the NLP area~\cite{gabor2018semeval}; 2) a multiple-domain corpus covering more abundant relations annotated on the publication abstracts from various artificial intelligence (AI) conference proceedings~\cite{luan2018multi}; and 3) the combination of previous two corpora where the distribution of data domains are unbalanced and annotations are provided by two different groups of annotators. The motivation of building this corpus is try to simulate the real data settings in digital libraries. Our ultimate goal is to help the stakeholders of digital libraries select the optimal tool to implement knowledge-based scientific information flows. 

In summary, we address the following research questions in this paper: 

\begin{enumerate}
    \item What is the impact of the eight classifiers on scientific relation classification?
    \item Which of the seven relation types studied are easy or challenging for classification?
    \item What is the practical relevance of the seven relation types in a scholarly knowledge graph?
\end{enumerate}

\section{Related Work}\label{sec:related}
\paragraph{\rm \textbf{Relations Mined from Scientific Publications.}} Overall, knowledge is organized in digital libraries based on the following three aspects of the digital collections: 1) metadata, 2) free-form content, and 3) ontologized content~\cite{dlko,kglib}. In this context, the main categories of relations that have been explored for scholarly publications belong to two groups. One group includes metadata relations such as authorship, co-authorship, and citations \cite{meta,coauthorship}. Research in this group mainly focuses on examining the social dimension of scholarly communication such as co-author prediction~\cite{coauthorship} and scholarly community analysis~\cite{meta}. The second group includes semantic relations, either as free-form semantic content classes~\cite{content,constituency} or as ontologized classes~\cite{ontology,scholarontology}. In the framework of automatic systems, content relations have been examined for: 1) scientific relation identification that involves determining which scientific term pairs are related~\cite{gabor2018semeval,constituency}, and 2) scientific relation classification that involves determining which relation type exists between related term pairs, where the relation types are typically pre-defined \cite{mre19,scibert,luan2018multi}. With respect to ontologized relation classes, prior work primarily considers the conceptual hierarchy based on formal concept analysis~\cite{ontology,scholarontology}.

We attempt the task of classifying semantic relations that were created from free-form text. Given that the digital libraries are interested in the creation of linked data~\cite{hallo2016current}, our attempted task directly facilitates the creation of scholarly knowledge graphs~\cite{auer2018towards}, offering structured data for use by librarians to generate linked data.

\paragraph{\rm \textbf{Techniques Developed for Relation Classification.}} Both rule-based~\cite{snowball} and learning-based~\cite{dependency,relrnn} methods have been developed for relation classification. Traditionally, learning-based systems relied on hand-crafted semantic and/or syntactic features~\cite{snowball,dependency}. In recent years, the success of deep learning techniques have nearly obviated the need to manually design features since they can more effectively learn latent feature representations for discriminating between relations. An attention-based bidirectional long short-term memory network (BiLSTM)~\cite{relrnn} was one of the first top-performing systems that leveraged neural attention mechanisms to capture important information per sentence for relation classification. Another advanced system~\cite{luan19} leveraged a dynamic span graph framework based on BiLSTMs to simultaneously extract terms and infer their pairwise relations. Aside from these neural methods considering the word sequence order, transformer-based models \cite{transformer} that use self-attention mechanisms to quantify the semantic association of each word to its context have become the current state-of-the-art in relation classification. E.g. \textsc{Bert} word embeddings~\cite{bert}. It can be trained to model data from any domain---the original \textsc{Bert} models were trained on books and Wikipedia. Now with the newly introduced \textsc{SciBert}~\cite{scibert}, there are \textsc{Bert} models trained on scholarly publications as well.

With respect to the classification strategy, the single-relation-at-a-time classification (SRC) that identifies the relation type for an entity pair each time are regularly adopted by prior work~\cite{relrnn,luan19,scibert}. To improve the classification efficiency, \cite{mre19} designed a \textsc{Bert}-based classifier that can recognize multiple pairwise relationships at one time, which can be regarded as a multiple-relations-at-a-time classification (MRC). Differing from prior work that emphasizes classification improvement, we focus on providing a fine-grained analysis of existing resources for selecting the proper tool to extract and organize scientific information in digital libraries.

\begin{table*}[tb]
\scriptsize
\begin{tabular}{|l|p{7.2cm}|c|c|c|c|c|c|}
\hline
\textbf{Id} & \textbf{Relation} & \multicolumn{2}{c}{\textbf{SemEval18}} & \multicolumn{2}{|c|}{\textbf{SciERC}} & \multicolumn{2}{c|}{\textbf{Combined}} \\
 & & Total & \% & Total & \% & Total & \% \\ \hline
1  & \textsc{Usage}: a scientific entity that is used for/by/on another scientific entity. E.g. \textit{MT system} is applied to \textit{Japanese} & 658 & 42.13\% & 2,437 & 52.43\% & 3,095 & 49.84\% \\ \hline
2  & \textsc{Feature-Of}: An entity is a characteristic or abstract model of another entity. E.g. \textit{computational complexity} of \textit{unification} & 392 & 25.10\% & 264 & 5.68\%  & 656 & 10.56\% \\ \hline
3  & \textsc{Conjunction}: Entities that are related in a lexical conjunction i.e., with `and' `or'. E.g. videos from \textit{Google Video} and a \textit{NatGeo documentary} & - & - & 582 & 12.52\%  & 582 & 9.37\% \\ \hline
4  & \textsc{Part-Of}: scientific entities that are in a part-whole relationship. E.g. describing the processing of \textit{utterances} in a \textit{discourse} & 304 & 19.46\% & 269 & 5.79\% & 573 & 9.23\% \\ \hline
5  & \textsc{Result}: An entity affects or yields a result. E.g. With only 12 \textit{training speakers} for SI recognition , we achieved a 7.5\% \textit{word error rate} & 92 & 5.89\% & 454 & 9.77\%  & 546 & 8.79\% \\ \hline
6  & \textsc{Hyponym-Of}: An entity whose semantic field is included within that of another entity. E.g. \textit{Image matching} is a problem in \textit{Computer Vision} & - & - & 409 & 8.80\%  & 409 & 6.59\%  \\ \hline
7 & \textsc{Compare}: An entity is compared to another entity. E.g.  \textit{conversation transcripts} have features that differ significantly from \textit{neat texts} & 116 & 7.43\% & 233 & 5.01\%  & 349 & 5.62\%   \\ \hline
\multicolumn{2}{|c|}{\textbf{Overall}} & 1,562 & 100\% & 4,648 & 100\%  & 6,210 & 100\% \\ \hline
\end{tabular}
\scriptsize
\caption{\footnotesize Overview of corpus statistics (also is accessible at \url{https://www.orkg.org/orkg/comparison/R38012}). `Total' and `\%' columns show the number and percentage of instances annotated with the corresponding relation over all abstracts, respectively.}
\vspace{-3ex}
\label{table:1}
\end{table*}

\section{Corpus}

In this study, we select two publicly available datasets~\cite{gabor2018semeval,luan2018multi} that contain a set of manually annotated scholarly abstracts for their scientific terms and the semantic relations between pairs of terms. Additionally, we combine these two datasets into a third new dataset, which offers a more realistic evaluation setting since it provides a larger, more diverse task representation. Table~\ref{table:1} shows the overall corpus statistics. The details of each corpus is provided next.

\paragraph{\rm \textbf{C1: The SemEval18 Corpus.}} This corpus was created for the seventh Shared Task organized at SemEval-2018~\cite{gabor2018semeval}. It comprised 500 abstracts of scholarly publications that are available at the ACL Anthology. Of these abstracts, 350 were partitioned as training data and the remaining 150 as testing data. In the abstracts, originally, six discrete semantic relations were defined to capture the predominant information content. For our evaluation, we omit one of the six, viz. \textsc{Topic}, since it is not well-represented in the corpus, and consider the following five relations: \textsc{Usage}, \textsc{Result}, \textsc{Model}, \textsc{Part\_Whole}, and \textsc{Comparison}. 

\paragraph{\rm \textbf{C2: The SciERC Corpus.}} Our second evaluation corpus~\cite{luan2018multi} also contains a set of 500 manually annotated abstracts of scholarly articles with their scientific terms and their pairwise relations. Unlike the SemEval18 corpus, the SciERC corpus represents diverse underlying data domains where the abstracts were taken from 12 AI conference/workshop proceedings in five research areas: artificial intelligence, natural language processing, speech, machine learning, and computer vision. These abstracts were annotated for the following seven relations: \textsc{Compare}, \textsc{Part-of}, \textsc{Conjunction}, \textsc{Evaluate-for}, \textsc{Feature-of}, \textsc{Used-for}, and \textsc{Hyponym-Of}. Similar to C1, this corpus was pre-partioned by the corpus creators. They adopted  a 350/50/100 train/development/testing dataset split. Comparing C2 with C1, we found that there are five relations except for \textsc{Conjunction} and \textsc{Hyponym-Of} in C2  that are semantically identical to the relations annotated in C1. 

\paragraph{\rm \textbf{C3: The Combined Corpus.}} 
Finally, this evaluation corpus was created by merging \textit{C1} and \textit{C2}. In the merging process, we renamed some relations that are semantically identical but have different labels. First, \textsc{Used-For} in \textit{C2} and \textsc{Usage} in \textit{C1} were unified as \textsc{Usage}. Further, by observing relation annotations in \textit{C1} and \textit{C2}, we found that \textsc{Result} in \textit{C1} and \textsc{Evaluate-For} in \textit{C2} essentially express the similar meaning but the arguments of these two relations were in the reverse order. For example, “[accuracy] for [semantic classification]” is labeled as ``accuracy" $\to$ \textsc{Evaluate-For} $\to$ ``semantic classification" in \textit{C2}, which can be regarded as ``semantic classification" $\to$ \textsc{Result} $\to$ ``accuracy". Therefore, we renamed all instances annotated with relation \textsc{Evaluate-For} in corpus \textit{C2} into \textsc{Result} by flipping their argument order. 
By combining 1000 abstracts with human annotations from two resources, our third evaluation corpus presents a comparatively more realistic evaluation scenario of large and heterogeneous data.

 
\section{\textsc{Bert}-based Scientific Relation Classifiers}

\textsc{Bert}~\cite{bert}, Bidirectional Encoder Representations from Transformers, as a pretrained language representation built on cutting-edge neural technology, provides NLP practitioners with high-quality language features from text data simply out-of-the-box that improves performance on many NLP tasks. 
These models return \textit{contextualized} embeddings for tokens which can be directly employed as features for various NLP tasks.
Further, with minimal task-specific extensions over the core \textsc{Bert} architecture, the embeddings can be relatively inexpensively fine-tuned to the task at hand, in turn facilitating even greater boosts in task performance. 

In this work, for scientific relation classification, we employ \textsc{Bert} embeddings and we also fine-tune them within two different neural extensions: 1) for single-relation-at-a-time classification (SRC); and 2) for multiple-relation-at-a-time classification (MRC). In the remainder of the section, we first describe the \textsc{Bert} models we employ followed by the SRC and MRC classifiers that implement different classification objectives.

\subsection{Pre-trained \textsc{Bert} Variants}
\textsc{Bert} models as pretrained language representations are available in several variants depending on the model configuration parameters and on the underlying training data. While there are over 16 types, in this work we select the following four core variants.


\paragraph{\textbf{\textsc{Bert}$_{\text{BASE}}$}}\footnote{https://github.com/google-research/bert} The first two models we use are in the category of the pretrained \textsc{Bert}$_{\text{BASE}}$. 
They were pretrained on billions of words from text data comprising the BooksCorpus (800M words)~\cite{zhu2015aligning} 
and English Wikipedia (2,500M words). The two models we select are: a cased model (where the case of the underlying words were preserved when training \textsc{Bert}$_{\text{BASE}}$) and an uncased model (where the underlying words were all lowercased when training \textsc{Bert}$_{\text{BASE}}$).


\paragraph{\textbf{\textsc{SciBert}}}\footnote{https://github.com/allenai/scibert} The next two models employed in this study are in the category of the pretrained scientific \textsc{Bert} called \textsc{SciBert}. They are language models based on \textsc{Bert} but trained on a large corpus of scientific text. 
Specifically, they are trained on a random sample of 1.14M papers from Semantic Scholar~\cite{ammar2018construction} consisting of full text of 18\% papers from the computer science domain and 82\% from the broad biomedical domain. 
Like \textsc{Bert}$_{\text{BASE}}$, for \textsc{SciBert}, we use its cased and uncased variants.

\subsection{Fine-tuned \textsc{Bert}-based Classifiers}
We implement the aforementioned \textsc{Bert} models within two neural system extensions that respectively adopt different classification strategies. 

\paragraph{\rm \textbf{Single-relation-at-a-time Classification (SRC)}} Classification models built for SRC generally extend the core \textsc{Bert} architecture with one additional linear classification layer that has $K \times H$ dimensions, where $K$ is the number of labels (i.e., relation types) and $H$ denotes the dimension of the word embedding space.  The label probabilities are further normalized by using a  softmax function and the classifier assigns the label with the maximum probability to each related concept pair. 

\paragraph{\rm \textbf{Multiple-relations-at-a-time Classification (MRC)}} 
This strategy is a more recent innovation on the classification problem in which the classifier can be trained with all the relation instances in a sentence at a time or predicts all the instances in one pass, as opposed to separately for each instance. In this case, however, the core \textsc{Bert} architecture's self-attention mechanism is modified to efficiently consider the representations of the relative positions of scientific terms~\cite{shaw2018self,mre19}. While this modification enables encoding the novel multiple-relations-at-a-time problem, for obtaining the classification probabilities, the MRC is also extended with a linear classification layer, though not identical to the SRC since it has to model the modified architecture. 

\section{Evaluation}

\subsection{Experimental Setup}

\paragraph{Experimental datasets, \textsc{Bert} word embeddings, and Classification strategies.} Our comprehensive evaluation setup involved three different corpora, four \textsc{Bert} embedding variants, and two classification strategies. Given this, we trained a total of \textit{eight} different classifiers, which for each of the three corpora resulted in 24 trained models.
Each corpus was already prepartitioned three ways as training/dev/test by the original creators, which we adopt. To train optimal classifiers on the respective corpus, we tuned the learning rate parameter $\eta$ for values \{2e-5, 3e-5, 5e-5\}. For other parameters such as the number of epochs, we used the default values in \textsc{SciBert} and \textsc{Bert} models.

\paragraph{Evaluation Metrics.} We employ the standard machine learning classification evaluation indicators, i.e., Precision ($P$), Recall ($R$), F1-score ($F1$), and Accuracy ($Acc$).


\begin{table}[!b]
\vspace{-3 ex}
\centering
\resizebox{1.0\textwidth}{!}{
    \begin{tabular}{l c c| c c| c c || c c| c c| c c ||c c }
    \toprule
\multirow{3}{*}{} & \multicolumn{6}{c ||}{\textbf{SRC}}                                                                   & \multicolumn{6}{c ||}{\textbf{MRC}}                                                                   & \multicolumn{2}{c}{\multirow{2}{*}{\textbf{Avg$\pm$Std}}} \\ \cmidrule{2-13}
                  & \multicolumn{2}{c |}{SemEval18} & \multicolumn{2}{c |}{SciERC} & \multicolumn{2}{c ||}{Combined} & \multicolumn{2}{c |}{SemEval18} & \multicolumn{2}{c |}{SciERC} & \multicolumn{2}{c ||}{Combined} & \multicolumn{2}{l}{}                           \\
                  & Acc.           & F1           & Acc.          & F1         & Acc.           & F1          & Acc.           & F1           & Acc.          & F1         & Acc.           & F1          & Acc.                    & F1                   \\ \hline \midrule
     Bert-base uncased & 76.42 & 71.74 & 84.6 & 77.25 & 81.75 & 77.38 & 80.4 & 79.98 & 83.42 & 74.84 & 80.84 & 76.29 & 81.24$\pm$2.84 & 76.25$\pm$2.78 \\  
    Bert-base cased & 73.58 & 71.14 & 85.32 & 77.92 & 78.73 & 74.38 & 79.55 & 78.44 & 83.72 & 75.07 & 79.42 & 74.8 & 80.05$\pm$4.14 & 75.29$\pm$2.65 \\ \midrule
    Scibert cased & 73.58 & 69.72 & \textbf{86.86} & \textbf{79.65} & \textbf{84.46} & \textbf{81.60} & 80.11 & 78.32 & 83.42 & 74.35 & \textbf{81.80} & \textbf{77.68} & 81.71$\pm$4.60 & 76.89$\pm$4.25\\ 
    Scibert uncased & \textbf{80.97} & \textbf{79.42} & 86.14 & 79.49  & 83.11 & 80.27 & \textbf{81.82} & \textbf{80.54}  & \textbf{84.33} & \textbf{77.44} & 81.06 & 76.76 & \textbf{82.91$\pm$2.04} & \textbf{78.99$\pm$1.54}\\ \midrule
    Avg. Scores & \multicolumn{6}{c ||}{Acc. 84.10   F1 80.22}& \multicolumn{6}{c||}{Acc. 82.35 F1 77.52}\\
    \bottomrule
    \end{tabular}
}
\scriptsize
\caption{Scientific relation classification results over three datasets (SemEval18, SciERC, \& Combined), four \textsc{Bert} model variants (\textsc{Bert} cased \& uncased; \textsc{SciBert} cased \& uncased), and two classification strategies (SRC \& MRC). \footnotesize{$Acc.$ is accuracy and $F1$ is the macro F1-score; Top scores are in bold.}}
\label{tab:overalleval}
\end{table}

\subsection{Results and Analysis}

In this section, we present results from our comprehensive evaluations with respect to the three main research questions that undergird this study.

\subsubsection{RQ1: What is the impact of the eight classifiers on scientific relation classification?}
The eight classifiers are obtained from two classification strategies built over four \textsc{Bert} model variants. We examine their classification results (depicted in Table~\ref{tab:overalleval}) in terms of the following three key characteristics of the classifiers.
\paragraph{The classification strategy, i.e., SRC vs. MRC.} From the $Acc$ and $F1$ shown in Table~\ref{tab:overalleval}, we see that SRC outperforms MRC on two corpora except the SemEval18 corpus. One characteristic of the SemEval18 corpus is that it has significantly lower number of annotations than the other two copora. Thus we infer that the novel MRC strategy is more robust than SRC because its performance level is unaffected by a drop in the number of the annotations.

\paragraph{Word embedding features, i.e., \textsc{Bert} vs. \textsc{SciBert}.} Regarding the \textsc{Bert} word embedding models, \textsc{SciBert} outperformed \textsc{Bert} on all three corpora with higher accuracy and F1 scores. Since our experimental corpora are all scholarly data, as an expected result, word embeddings trained on the similar data domains are better suited. 

\paragraph{Vocabulary case in \textsc{Bert} models, i.e., cased vs. uncased.} We observe that the uncased \textsc{Bert} models (\textsc{SciBert}: 82.91, \textsc{Bert}: 81.24) show higher classification accuracy than their cased counterpart (\textsc{SciBert}: 81.71, \textsc{Bert}: 80.05) on average. Further, the uncased models have a lower standard deviation in accuracy overall (\textsc{SciBert}: 2.04, \textsc{Bert}: 2.84) than the cased models (\textsc{SciBert}: 4.60, \textsc{Bert}: 4.14); comparisons on $F1$ are along similar lines. Hence, our results suggest that uncased \textsc{Bert} models can achieve more stable performances than cased variants.

In conclusion, with respect to the classification strategy, we find SRC outperforms MRC (see averaged scores in the last row in Table~\ref{tab:overalleval}). Nevertheless, the advanced MRC strategy demonstrates consistently robust performance that remains relatively unaffected by smaller dataset sizes compared to the SRC (e.g. SRC vs. MRC results on the SemEval18 corpus). On the other hand, with respect to \textsc{Bert} word embedding variants, from the averaged scores in the last column in Table~\ref{tab:overalleval}, the \textsc{SciBert} uncased model posits as the optimal word embedding features model on scholarly articles. 

\begin{table}[!b]
\centering
    \begin{tabular}{l@{\hskip 0.8cm} c@{\hskip 0.2cm} c@{\hskip 0.2cm} c@{\hskip 0.2cm} c@{\hskip 0.2cm} c@{\hskip 0.2cm} c }
    \toprule
    \multirow{2}{*}{\thead{Relationship Type\\SemEval18}} & \multicolumn{3}{c}{\textbf{SRC}} & \multicolumn{3}{c}{\textbf{MRC}} \\ \cmidrule(lr){2-4} \cmidrule(lr){5-7} 
    & P & R & F1 & P & R & F1     \\ \hline \midrule
    \textsc{Usage} & \textbf{87.22} & 89.71 & \textbf{88.45} & 90.53 & \textbf{87.43} & \textbf{88.95} \\
    \textsc{Result} & 78.26 & \textbf{90.00} & 83.72 & \textbf{100.00} & 75.00 & 85.71 \\
    \textsc{Compare} & 85.71 & 85.71 & 85.71 & 75.00 & 85.71 & 80.00 \\
    \textsc{Model-Feature} & 66.67 & 75.76 & 70.92 & 70.83 & 77.27 & 73.91 \\
    \textsc{Part-Whole} & 79.25 & 60.00 & 68.29 & 70.83 & 72.86 & 71.83 \\
    \bottomrule
    \end{tabular}
\caption{Per-relation classification scores of SRC and MRC best systems on SemEval18.}
\label{tab:relsemeval}
\end{table}

\subsubsection{RQ2: Which of the seven relation types are easy/challenging to be classified?}
Examining the fine-grained per-relation classification results in Tables~\ref{tab:relsemeval} to~\ref{tab:relcomb} across all our evaluation corpora for both SRC and MRC, we note the classification ranked order. 
Of all relations, \textsc{Usage} (\textsc{Used-For}) is the easiest classification target, since in all three tables, it is in the two topmost. \textsc{Usage} is the most predominant type in all corpora. This accounts, in part, for its high-scored classification, since the classifiers are trained on a significant number of training instances for \textsc{Usage} compared to the rest.

\begin{table}[!tb]
\centering
    \begin{tabular}{l@{\hskip 0.8cm} c@{\hskip 0.2cm} c@{\hskip 0.2cm} c@{\hskip 0.2cm} c@{\hskip 0.2cm} c@{\hskip 0.2cm} c }
    \toprule
    \multirow{2}{*}{\thead{Relationship Type\\SciERC}} & \multicolumn{3}{c}{\textbf{SRC}} & \multicolumn{3}{c}{\textbf{MRC}} \\ \cmidrule(lr){2-4} \cmidrule(lr){5-7} 
     & P & R & F1 & P & R & F1     \\ \hline \midrule
    \textsc{Used-For} & \textbf{93.30} & 91.37 & \textbf{92.32} & \textbf{88.75} & 90.24 & \textbf{89.49} \\
    \textsc{Conjunction} & 87.97 & \textbf{95.12} & 91.41 & 80.69 & \textbf{95.12} & 87.31 \\
    \textsc{Hyponym-Of} & 92.31 & 89.55 & 90.91 & 80.00 & 82.93 & 81.44\\
    \textsc{Evaluate-For} & 82.29 & 86.81 & 84.49 & 84.44 & 83.52 & 83.98 \\
    \textsc{Compare} & 72.73 & 84.21 & 78.05 & 83.87 & 68.42 & 75.36\\
    \textsc{Part-Of} & 66.04 & 55.56 & 60.34 & 65.52 & 60.32 & 62.81\\
    \textsc{Feature-Of} & 59.02 & 61.02 & 60.00 & 73.68 & 47.46 & 57.73\\
    \bottomrule
    \end{tabular}
\caption{Per-relation classification results of SRC and MRC best systems on SciERC.}
\label{tab:relscierc}
\vspace{-3ex}
\end{table}

\begin{table}[!tb]
\centering
    \begin{tabular}{l@{\hskip 0.8cm} c@{\hskip 0.2cm} c@{\hskip 0.2cm} c@{\hskip 0.2cm} c@{\hskip 0.2cm} c@{\hskip 0.2cm} c }
    \toprule
    \multirow{2}{*}{\thead{Relationship Type\\Combined}} & \multicolumn{3}{c}{\textbf{SRC}} & \multicolumn{3}{c}{\textbf{MRC}} \\ \cmidrule(lr){2-4} \cmidrule(lr){5-7} 
    & P & R & F1 & P & R & F1     \\ \hline \midrule
    \textsc{Conjunction} & \textbf{92.56} & \textbf{91.06} & \textbf{91.80} & 85.07 & \textbf{92.68} & \textbf{88.72}\\
    \textsc{Usage}  & 91.30 & 88.98 & 90.13 & \textbf{87.96} & 87.71 & 87.84\\
    \textsc{Hyponym-Of}  & 89.39 & 88.06 & 88.72 & 83.12 & 78.05 & 80.50\\
    \textsc{Compare}  & 86.89 & 89.83 & 88.33 & 73.85 & 81.36 & 77.41\\
    \textsc{Result} & 76.36 & 75.68 & 76.02 & 84.69 & 74.77 & 79.43\\
    \textsc{Part-Of} & 75.86 & 66.17 & 70.68 & 68.33 & 61.65 & 64.82\\
    \textsc{Feature-Of}  & 58.02 & 75.20 & 65.51 & 60.28 & 68.00 & 63.91\\
    \bottomrule
    \end{tabular}
\scriptsize
\caption{Per-relation classification results of the best SRC and MRC systems on the Combined corpus.}
\label{tab:relcomb}
\end{table}

For the challenging relations, we examine the results per corpus. Starting with the Table~\ref{tab:relsemeval} results for the SemEval18 corpus, we observe that both SRC and MRC find \textsc{Part-Whole} most challenging to classify. We surmise that this relation displays high diversity in the underlying natural language text from which it is induced; hence the classifier is unable to generalize a consistent set of patterns for it. Observing classification performance ranks, for three of the five relations (i.e., \textsc{Usage}, \textsc{Model-Feature} and \textsc{Part-Whole}), SRC and MRC obtain the same classification rank order. For \textsc{Result} and \textsc{Compare}, they are opposites, where SRC classifies \textsc{Result} better than MRC.

 
In Table~\ref{tab:relscierc} results for SciERC, both classifiers perform significantly low on two relations, viz. \textsc{Feature-Of} and \textsc{Part-Of}. Since these two relations are not the most underrepresented in the corpus, we theorize that their low classification performance is owed to the natural language text diversity from which they are deduced. In this case, obtaining more annotated instances is one way to boost classifier performance. In terms of the ranked order of performances on the relations, SRC and MRC perform identically on SciERC data. 

And lastly in Table~\ref{tab:relcomb} results on the Combined corpus, for the challenging relations, both SRC and MRC have the same result as they did on SciERC---i.e., \textsc{Feature-Of} followed by \textsc{Part-Of} are the most challenging. And we theorize the same reason for the low scores on these relations, since Combined contains SciERC data. Given the two corpora in the Combined dataset, SciERC additionally introduced \textsc{Conjunction} which SemEval18 did not have. \textsc{Conjunction} is among the top two easiest relations to classify, with \textsc{Usage} as the other, for the classifiers trained on SciERC and on the Combined corpus. Further, its classification is better in the Combined corpus than in SciERC. This lends an understanding to the realistic evaluation settings that the Combined corpus presents. To elaborate, for \textsc{Usage}, instances from SemEval18 and SciERC (i.e. \textsc{Used-For}) are combined, resulting in an insignificant dip in performance (on the Combined corpus, \textsc{Usage} ranks second easiest compared with SemEval18 and SciERC) since they are now non-uniform annotation signals. As opposed to the case of \textsc{Conjunction}, the Combined corpus obtains a uniform annotation signal from just the SciERC corpus and ranks a minor degree higher at classifying it.

Finally, a list summarizing the top-scoring per-relation performances for scientific relation classification across all three tables, includes the following: \textsc{Usage} (SRC in SciERC), \textsc{Conjunction} (SRC in Combined), \textsc{Hyponym-Of} (SRC in SciERC), \textsc{Result} (MRC in SemEval18), \textsc{Part-Of} (MRC in SemEval18 for \textsc{Part-Whole}), and \textsc{Feature-Of} (MRC in SemEval18 for \textsc{Model-Feature}). Since the SemEval18 corpus appears the most times in the top-ranked results, we conclude that its annotations obtain a relatively better trained classifier. However, the SemEval18 corpus only includes scholarly abstracts from one AI domain i.e. NLP (in the ACL Anthology), whereas SciERC is more comprehensively inclusive across various AI domains. Thus, an additional factor that classifiers trained on SciERC handle is domain diversity. 

\begin{figure*}[t]
   \begin{minipage}{0.50\textwidth}
     \centering
     \includegraphics[scale = 0.25]{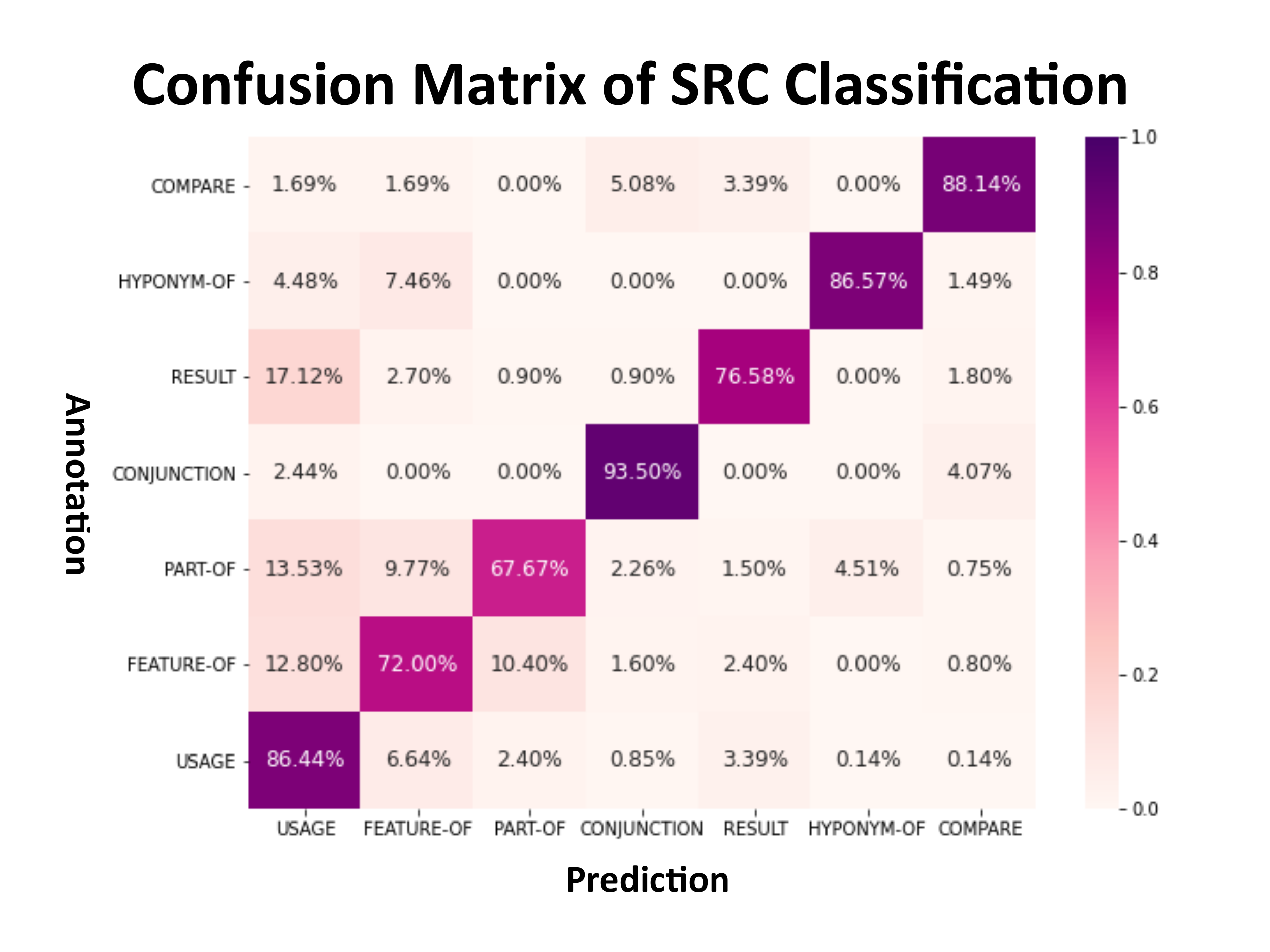}
   \end{minipage}\hfill
   \begin{minipage}{0.50\textwidth}
     \centering
     \includegraphics[scale = 0.25]{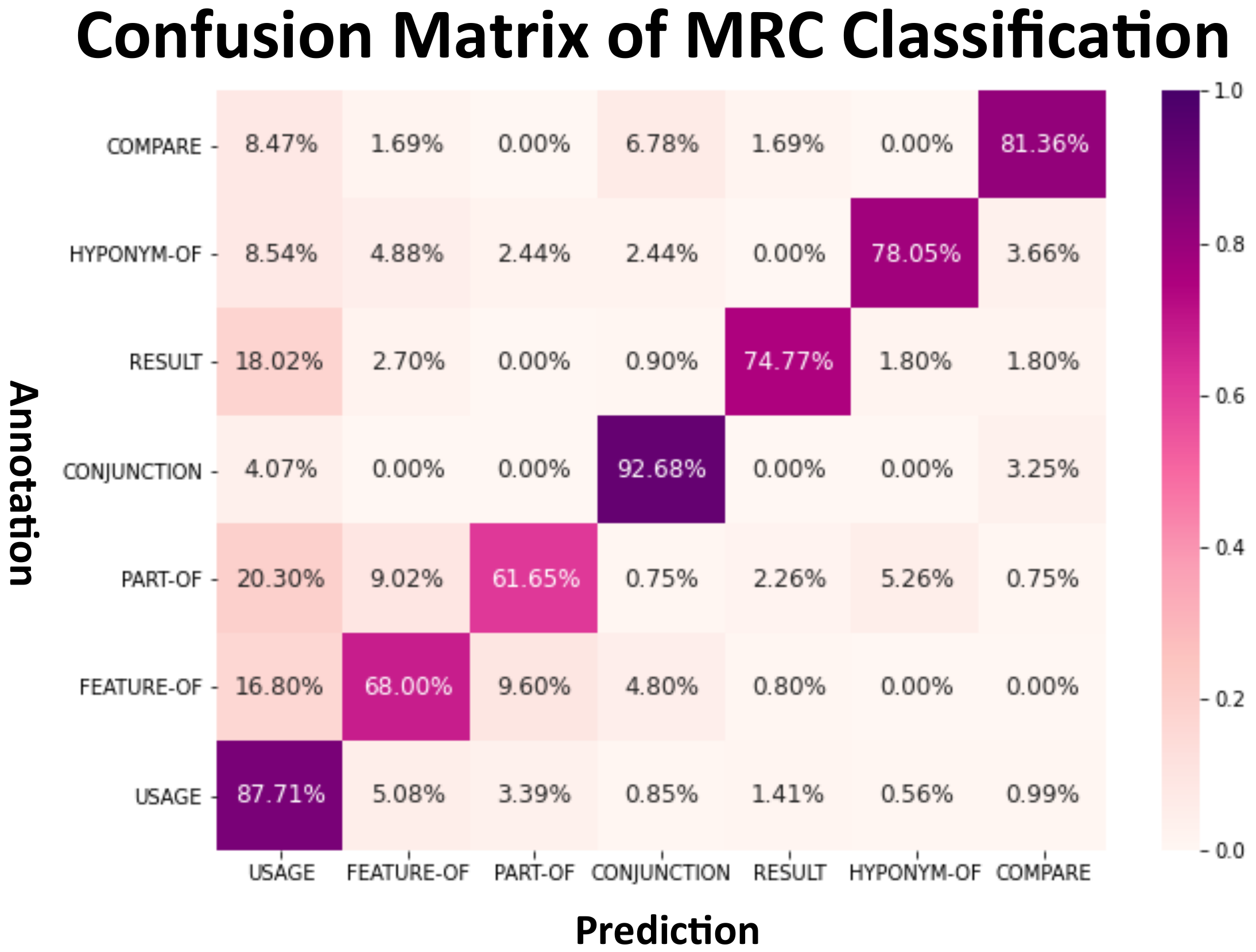}
   \end{minipage}
   \caption{Confusion matrices for SRC and MRC misclassifications on the Combined data. 
   }
   \label{fig:conmat}
\end{figure*}

\begin{figure*}[b]
    \centering
    \vspace{-2 ex}
    \includegraphics[width =0.75\textwidth, height = 4.0cm]{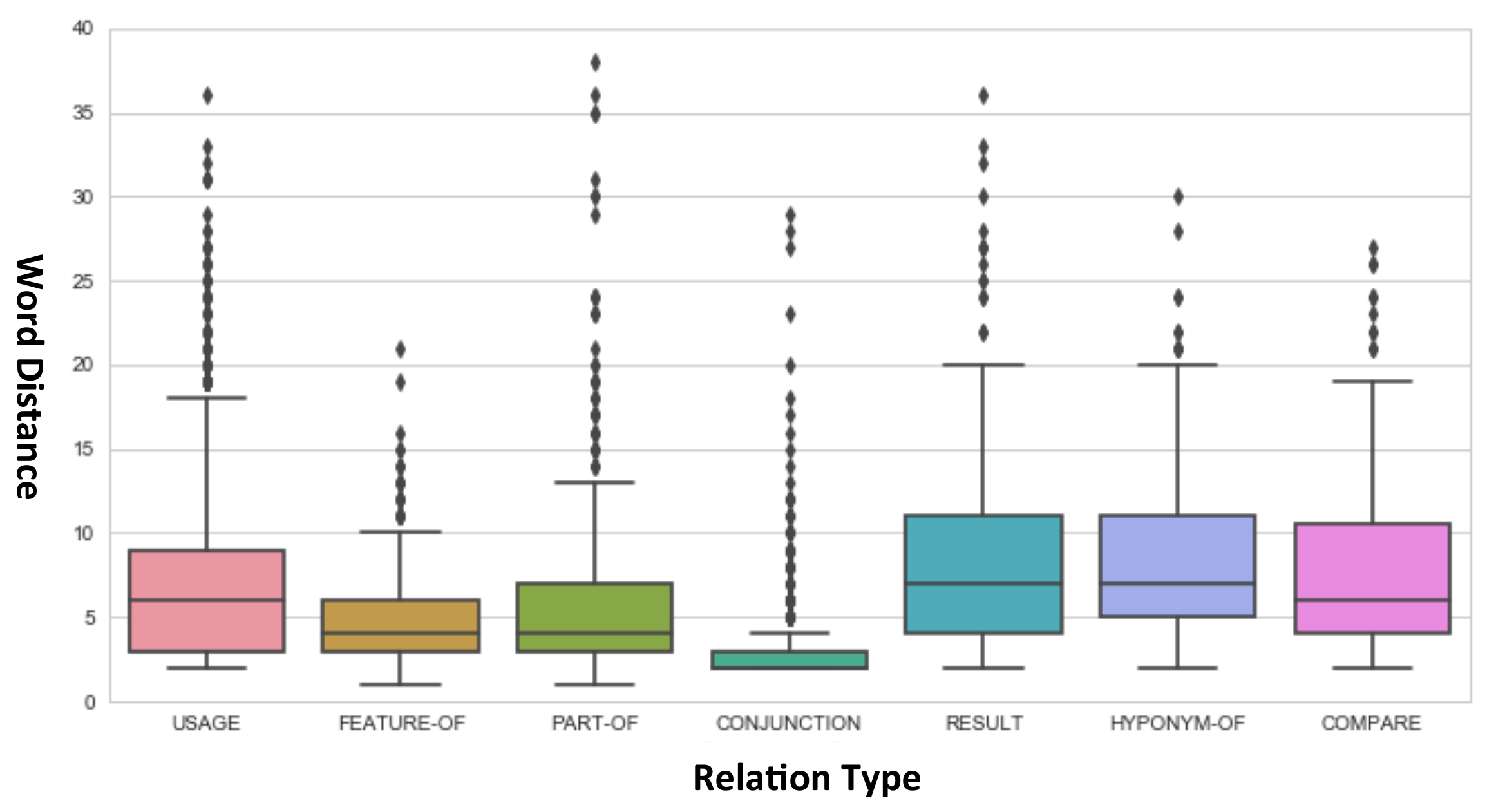}
    \caption{Distributions of word distances in the combined corpus text between scientific term pairs.}
    \label{fig:worddist}
\end{figure*}

\subsubsection{Error Analysis} A closer look at the misclassifications is portrayed in the confusion matrices in Figure~\ref{fig:conmat} for the SRC and MRC strategies on the Combined corpus. Four of the seven relations, i.e. \textsc{Hyponym-Of}, \textsc{Result}, \textsc{Part-Of}, and \textsc{Feature-Of}, are highly likely to be misclassified as \textsc{Usage}. This shows that our classifiers are biased by the predominant \textsc{Usage} relation. In general, unbalanced distribution of training samples (see the details in the corpus section) is, more often than not, one of the main factors for confusion learned in machine learning systems. For the most challenging relations \textsc{Feature-Of} and \textsc{Part-Of}, after \textsc{Usage}, are highly likely to be confused with each other (\textsc{Feature-Of} as \textsc{Part-Of} ($\sim$10\% confusion), and vice-versa ($\sim$9.4\% confusion)). For the relations \textsc{Hyponym-Of} and \textsc{Feature-Of} that loosely demonstrate a relation heirarchy such that \textsc{Hyponym-Of} subsumes \textsc{Feature-Of}, but not the other way around, we find the classification confusion demonstrates a consistent pattern to this data. From the matrices, we see that \textsc{Hyponym-Of} has $\sim$6\% likelihood to be predicted as \textsc{Feature-Of}, but none of the \textsc{Feature-Of} (0\%) instances were confused with \textsc{Hyponym-Of}. 

To offer another pertinant angle on the classifier error analysis, we compute the distribution of word distances between related scientific term pairs in the Combined corpus. The result is depicted in Figure~\ref{fig:worddist}. In general, the majority of box plots shown in Figure~\ref{fig:worddist} are skewed with a long upper whisker and a short lower whisker. This pattern indicates that the distance between paired scientific terms is typically closed in the text. Differ from other relations, the word distance of \textsc{CONJUNCTION} is much shorter, which makes sense because term pairs with this relationship are typically connected by a single connection term such as ``and" and ``or". This consistent pattern could be another reason why \textsc{Conjunction} is comparatively easier than other relations to be classified. Further, the average word distance of \textsc{Feature-Of}, \textsc{Part-Of}, \textsc{Hyponym-Of}, and \textsc{Compare} is closer to the lower quartile than the other relations. Such varied distribution may bring challenges for a classifier to identify these relations. Notably, the similar median value and spread range between \textsc{Feature-Of} and \textsc{Part-Of} could account for why they are challenging to be identified by the classification models.

\subsubsection{RQ3: What is the practical relevance of the seven relations studied in this paper in a scholarly knowledge graph?} 
\begin{figure*}[t]
    \centering
    \includegraphics[width =0.9\textwidth]{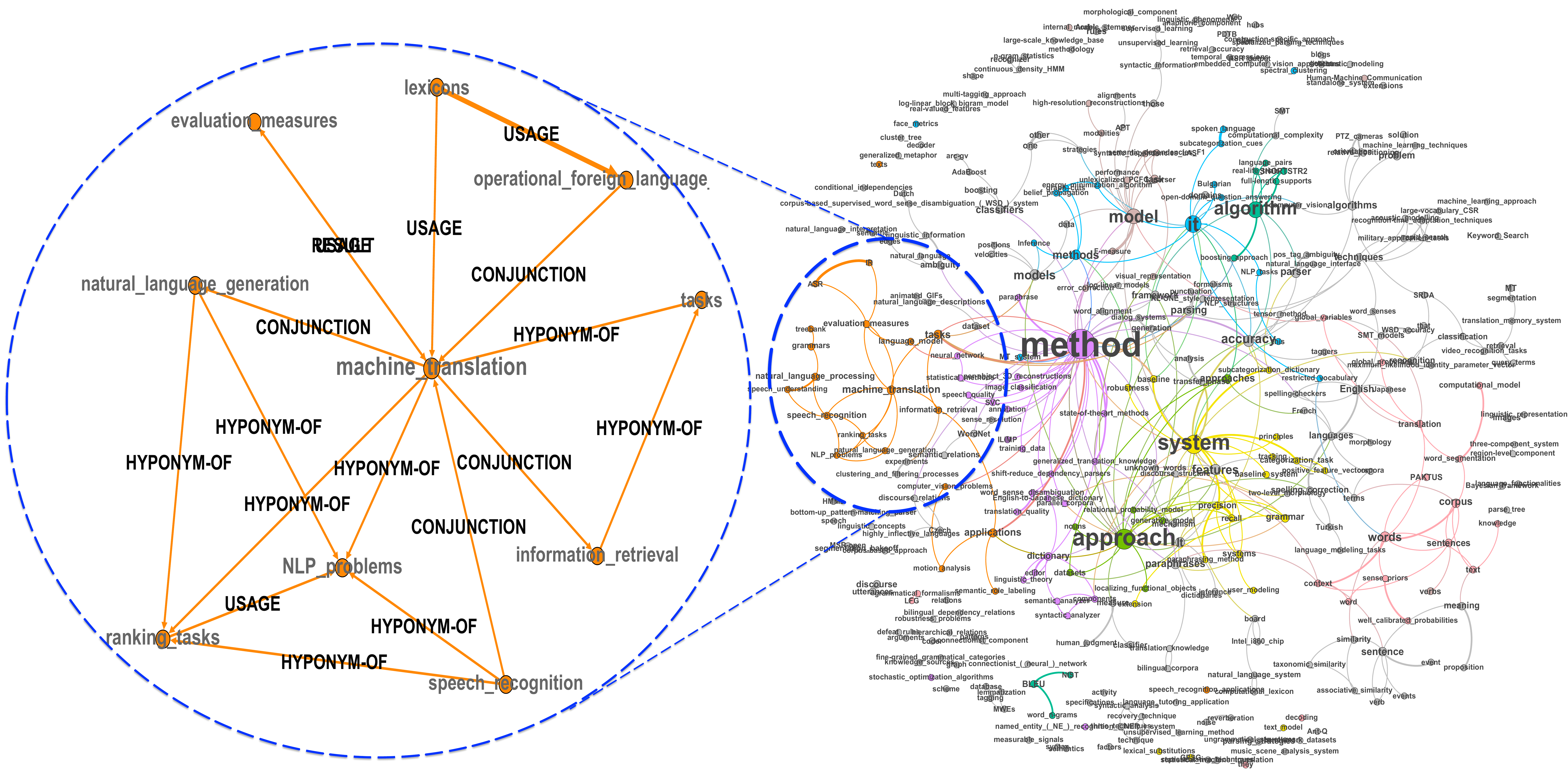}
    \scriptsize
    \caption{A knowledge graph constructed from the relation triples in the Combined corpus. The graph on the left is the ``ego-network'' of the scientific term ``machine\_translation''. The node size is determined by node weighted degree. Colors denote the modularity classes based on the graph structure. \scriptsize{The graph was generated using https://gephi.org/}}
    \label{fig:kg}
    \vspace{-1 ex}
\end{figure*}

As a practical illustration of the relation triples studied in this work, we build a knowledge graph from their annotations in the 1000 scholarly abstracts in the Combined dataset. This is depicted in Figure~\ref{fig:kg}. Looking at the corpus-level graph (the right graph), we observe that generic scientific terms such as ``method,'' ``approach,'' and ``system'' are the most densely connected nodes, as expected since generic terms are found across research areas. In the zoomed-in ego-network of the term ``machine\_translation'' (the left graph), \textsc{Hyponym-Of} is meaningfully highlighted by its role linking ``machine\_translation'' and its sibling nodes as the research tasks ``speech\_recognition,'' and ``natural\_language\_generation'' to the parent node ``NLP\_problems.'' The term ``lexicon'' is related by \textsc{Usage} to ``machine\_translation'' and ``operational\_foreign\_language.'' The \textsc{Conjunction} link joins ``machine\_translation'' and ``speech\_recognization'' both aim at translating information from one source to the other one. This knowledge graph now enables various property comparisons across 1000 scholarly abstracts (consider the corpus statistics Table 1 generated from the Open Research Knowledge Graph presented in the corpus section).

\section{Conclusions and Recommendations}
We have investigated the scientific relation classification task for improving scholarly knowledge representations in digital libraries. Our surveyed systems offer a comprehensive view of eight \textsc{Bert}-based classification models. Our observations indicate that the performance of classifiers are mainly associated with two aspects. First, from the perspective of training algorithms, three main factors including classification strategies, the pre-training corpus domain and vocabulary case are important to be considered when selecting the appropriate model to apply in practice. Second, with respect to the annotations for training, for each predefined relation, the labeling size and the grammatical patterns in language expressions are critical facets that determine the ability of a \textsc{Bert}-based classification model to identify this relation type. 

In summary, We provide the following recommendations to the stakeholders of digital libraries for applying the optimal technique to automatically classify scientific relations from scholarly articles: 
\begin{itemize}
    \item Domain-specific pre-training corpus benefits the performance of \textsc{Bert}-based classifiers.
    \item According to the classification strategy, although SRC outperforms MRC in general, the latter strategy demonstrates consistently robust performance on the small training corpus.
    \item Overall, the uncased \textsc{Bert} models achieves better and more stable performance than the cased models. 
    \item For each pre-defined relation, the large annotation size and/or fixed syntactic structure in expressions helps the model to learn the latent features of this relation during training.
\end{itemize}


\section{Future Work}
To further facilitate the choice of the proper technique for classifying scientific relations toward the creation of structured, semantic representations over scholarly articles, there are three main avenues that are worthwhile for future exploration. As we have seen in the process of examining our \textbf{RQ2}, there exist label biases in the annotated corpora such that some relations are better represented than others (E.g. \textsc{Usage}). Toward this end, such data needs to be further curated by experts to enable a well-represented domain. Further, digital libraries deal with various domains in Science in general. While our evaluations have been performed on corpus that covers the Artificial Intelligence research area, there still remains a plenty of potentials to explore other research domains that are unrelated to Computer Science specifically. Finally, we have examined scientific relation classification in terms of seven relations, ontologized models of the scientific world~\cite{fathalla2017towards,scholarontology} posit a larger set of relations or properties. For this, techniques such as open information extraction or ontology-based extraction are viable alternatives for future developments.

\bibliographystyle{splncs04}
\bibliography{tpdl}

\begin{thebibliography}{10}
\providecommand{\url}[1]{\texttt{#1}}
\providecommand{\urlprefix}{URL }
\providecommand{\doi}[1]{https://doi.org/#1}

\bibitem{snowball}
Agichtein, E., Gravano, L.: Snowball: Extracting relations from large
  plain-text collections. In: Proceedings of the fifth ACM conference on
  Digital libraries. pp. 85--94 (2000)

\bibitem{ammar2018construction}
Ammar, W., Groeneveld, D., Bhagavatula, C., Beltagy, I., Crawford, M., Downey,
  D., Dunkelberger, J., Elgohary, A., Feldman, S., Ha, V., et~al.: Construction
  of the literature graph in semantic scholar. In: NAACL, Volume 3 (Industry
  Papers). pp. 84--91 (2018)

\bibitem{auer2018towards}
Auer, S., Kovtun, V., Prinz, M., Kasprzik, A., Stocker, M., Vidal, M.E.:
  Towards a knowledge graph for science. In: Proceedings of the 8th
  International Conference on Web Intelligence, Mining and Semantics. pp.~1--6
  (2018)

\bibitem{auer2019orkg}
Auer, S., Mann, S.: Toward an open knowledge research graph. The Serials
  Librarian  \textbf{76} (2019)

\bibitem{augenstein2017semeval}
Augenstein, I., Das, M., Riedel, S., Vikraman, L., McCallum, A.: Semeval 2017
  task 10: Scienceie-extracting keyphrases and relations from scientific
  publications. In: Proceedings of SemEval-2017. pp. 546--555 (2017)

\bibitem{scibert}
Beltagy, I., Lo, K., Cohan, A.: {S}ci{BERT}: A pretrained language model for
  scientific text. In: EMNLP-IJCNLP. pp. 3615--3620. ACL, Hong Kong, China (Nov
  2019)

\bibitem{dependency}
Culotta, A., Sorensen, J.: Dependency tree kernels for relation extraction. In:
  Proceedings of the 42nd annual meeting on ACL. p.~423. ACL (2004)

\bibitem{bert}
Devlin, J., Chang, M.W., Lee, K., Toutanova, K.: {BERT}: Pre-training of deep
  bidirectional transformers for language understanding. In: Proceedings of
  NAACL, Volume 1 (Long and Short Papers). pp. 4171--4186. ACL, Minneapolis,
  Minnesota (Jun 2019)

\bibitem{fathalla2017towards}
Fathalla, S., Vahdati, S., Auer, S., Lange, C.: Towards a knowledge graph
  representing research findings by semantifying survey articles. In: TPDL. pp.
  315--327. Springer (2017)

\bibitem{gabor2018semeval}
G{\'a}bor, K., Buscaldi, D., Schumann, A.K., QasemiZadeh, B., Zargayouna, H.,
  Charnois, T.: Semeval-2018 task 7: Semantic relation extraction and
  classification in scientific papers. In: Proceedings of SemEval-2018. pp.
  679--688 (2018)

\bibitem{hallo2016current}
Hallo, M., Luj{\'a}n-Mora, S., Mat{\'e}, A., Trujillo, J.: Current state of
  linked data in digital libraries. Journal of Information Science
  \textbf{42}(2),  117--127 (2016)

\bibitem{kglib}
Haslhofer, B., Isaac, A., Simon, R.: Knowledge graphs in the libraries and
  digital humanities domain. arXiv preprint arXiv:1803.03198  (2018)

\bibitem{Jaradeh2019ORKG}
Jaradeh, M.Y., Oelen, A., Farfar, K.E., Prinz, M., D'Souza, J., Kismih\'{o}k,
  G., Stocker, M., Auer, S.: Open research knowledge graph: Next generation
  infrastructure for semantic scholarly knowledge. In: KCAP. pp. 243--246. ACM,
  New York, NY, USA (2019)

\bibitem{constituency}
Jiang, M., Diesner, J.: A constituency parsing tree based method for relation
  extraction from abstracts of scholarly publications. In: Proceedings of the
  Thirteenth Workshop on Graph-Based Methods for Natural Language Processing
  (TextGraphs-13). pp. 186--191 (2019)

\bibitem{content}
Klampfl, S., Kern, R.: An unsupervised machine learning approach to body text
  and table of contents extraction from digital scientific articles. In: TPDL.
  pp. 144--155. Springer (2013)

\bibitem{luan2018multi}
Luan, Y., He, L., Ostendorf, M., Hajishirzi, H.: Multi-task identification of
  entities, relations, and coreference for scientific knowledge graph
  construction. In: EMNLP. pp. 3219--3232 (2018)

\bibitem{luan19}
Luan, Y., Wadden, D., He, L., Shah, A., Ostendorf, M., Hajishirzi, H.: A
  general framework for information extraction using dynamic span graphs. arXiv
  preprint arXiv:1904.03296  (2019)

\bibitem{manning2015computational}
Manning, C.D.: Computational linguistics and deep learning. Computational
  Linguistics  \textbf{41}(4),  701--707 (2015)

\bibitem{scholarontology}
Quan, T.T., Hui, S.C., Fong, A.C.M., Cao, T.H.: Automatic generation of
  ontology for scholarly semantic web. In: ISWC. pp. 726--740. Springer (2004)

\bibitem{shaw2018self}
Shaw, P., Uszkoreit, J., Vaswani, A.: Self-attention with relative position
  representations. In: NAACL-HLT (2) (2018)

\bibitem{ontology}
Silvescu, A., Reinoso-Castillo, J., Honavar, V.: Ontology-driven information
  extraction and knowledge acquisition from heterogeneous, distributed,
  autonomous biological data sources. In: In Proceedings of the IJCAI-2001
  Workshop on Knowledge Discovery from Heterogeneous, Distributed, Autonomous,
  Dynamic Data and Knowledge Sources (2001)

\bibitem{coauthorship}
Sivasubramaniam, A., Debnath, S., Li, H., Lee, W.C., Bolelli, L., Giles, C.L.,
  Zhuang, Z., Councill, I.G.: Learning metadata from the evidence in an on-line
  citation matching scheme. In: Proceedings of JCDL'06. pp. 276--285. IEEE
  (2006)

\bibitem{dlko}
Soergel, D.: Digital libraries and knowledge organization. In: Semantic digital
  libraries, pp. 9--39. Springer (2009)

\bibitem{meta}
Vahdati, S., Palma, G., Nath, R.J., Lange, C., Auer, S., Vidal, M.E.: Unveiling
  scholarly communities over knowledge graphs. In: TPDL. pp. 103--115. Springer
  (2018)

\bibitem{transformer}
Vaswani, A., Shazeer, N., Parmar, N., Uszkoreit, J., Jones, L., Gomez, A.N.,
  Kaiser, {\L}., Polosukhin, I.: Attention is all you need. In: Advances in
  NIPS. pp. 5998--6008 (2017)

\bibitem{mre19}
Wang, H., Tan, M., Yu, M., Chang, S., Wang, D., Xu, K., Guo, X., Potdar, S.:
  Extracting multiple-relations in one-pass with pre-trained transformers. In:
  Proceedings of the 57th Annual Meeting of the ACL. pp. 1371--1377. ACL,
  Florence, Italy (Jul 2019)

\bibitem{stm}
Ware, M., Mabe, M.: The stm report: An overview of scientific and scholarly
  journal publishing  (03 2015)

\bibitem{stephen}
Weigl, D.M., Kudeki, D.E., Cole, T.W., Downie, J.S., Jett, J., Page, K.R.:
  Combine or connect: Practical experiences querying library linked data.
  Proceedings of the Association for Information Science and Technology
  \textbf{56}(1),  296--305 (2019)

\bibitem{relrnn}
Zhou, P., Shi, W., Tian, J., Qi, Z., Li, B., Hao, H., Xu, B.: Attention-based
  bidirectional long short-term memory networks for relation classification.
  In: Proceedings of the 54th annual meeting of the ACL (volume 2: Short
  papers). pp. 207--212 (2016)

\bibitem{zhu2015aligning}
Zhu, Y., Kiros, R., Zemel, R., Salakhutdinov, R., Urtasun, R., Torralba, A.,
  Fidler, S.: Aligning books and movies: Towards story-like visual explanations
  by watching movies and reading books. In: Proceedings of the IEEE
  international conference on computer vision. pp. 19--27 (2015)

\end{thebibliography}

\end{document}